\documentclass[12pt]{article}

\begin{document}

\title{Whence the Gauge Fields arise}
\author{G. Quznetsov \\
gunn@yahoo.com}
\maketitle

\begin{abstract}
A gauge fields (and massive, too) arise from the production of the
probability by the spinors.
\end{abstract}

I use the following notation:

\[
1_2=\left[ 
\begin{array}{cc}
1 & 0 \\ 
0 & 1
\end{array}
\right] \mbox{,}0_2=\left[ 
\begin{array}{cc}
0 & 0 \\ 
0 & 0
\end{array}
\right] \mbox{, } 
\]

\[
\sigma _x=\left[ 
\begin{array}{cc}
0 & 1 \\ 
1 & 0
\end{array}
\right] \mbox{, }\sigma _y=\left[ 
\begin{array}{cc}
0 & -i \\ 
i & 0
\end{array}
\right] \mbox{, }\sigma _z=\left[ 
\begin{array}{cc}
1 & 0 \\ 
0 & -1
\end{array}
\right] , 
\]

\[
\beta _1=\left[ 
\begin{array}{cc}
\sigma _x & 0_2 \\ 
0_2 & -\sigma _x
\end{array}
\right] \mbox{, }\beta _2=\left[ 
\begin{array}{cc}
\sigma _y & 0_2 \\ 
0_2 & -\sigma _y
\end{array}
\right] \mbox{, }\beta _3=\left[ 
\begin{array}{cc}
\sigma _z & 0_2 \\ 
0_2 & -\sigma _z
\end{array}
\right] , 
\]

\[
\beta _4=i\cdot \left[ 
\begin{array}{cc}
0_2 & 1_2 \\ 
-1_2 & 0_2
\end{array}
\right] \mbox{, }\beta _0=\left[ 
\begin{array}{cc}
1_2 & 0_2 \\ 
0_2 & 1_2
\end{array}
\right] =1_4\mbox{, }\gamma _0=\left[ 
\begin{array}{cc}
0_2 & 1_2 \\ 
1_2 & 0_2
\end{array}
\right] =\beta _5\mbox{,} 
\]

\[
\gamma _5=\left[ 
\begin{array}{cc}
1_2 & 0_2 \\ 
0_2 & -1_2
\end{array}
\right] \mbox{,}0_4=\left[ 
\begin{array}{cc}
0_2 & 0_2 \\ 
0_2 & 0_2
\end{array}
\right] \mbox{, } 
\]

Here $\beta _1$, $\beta _2$, $\beta _3$, $\beta _4$, $\gamma _0$ are all
five members of the light Clifford pentad \cite{MD},\cite{GQ} and

\[
\gamma _5=i\cdot \gamma _0\cdot \beta _4\mbox{,} 
\]

\section{The probability current vector}

Let

\[
\left\langle \rho ,j_x,j_y,j_z\right\rangle 
\]

be a probability current vector and $\psi $ be any complex 4-spinor:

\[
\psi =\left| \psi \right| \cdot \left[ 
\begin{array}{c}
\exp \left( i\cdot g\right) \cdot \cos \left( b\right) \cdot \cos \left(
a\right) \\ 
\exp \left( i\cdot d\right) \cdot \sin \left( b\right) \cdot \cos \left(
a\right) \\ 
\exp \left( i\cdot f\right) \cdot \cos \left( v\right) \cdot \sin \left(
a\right) \\ 
\exp \left( i\cdot q\right) \cdot \sin \left( v\right) \cdot \sin \left(
a\right)
\end{array}
\right] \mbox{.} 
\]

In this case the following system of equations

\begin{equation}
\left\{ 
\begin{array}{c}
\psi ^{\dagger }\cdot \psi =\rho \mbox{,} \\ 
\psi ^{\dagger }\cdot \beta _1\cdot \psi =j_x\mbox{,} \\ 
\psi ^{\dagger }\cdot \beta _2\cdot \psi =j_y\mbox{,} \\ 
\psi ^{\dagger }\cdot \beta _3\cdot \psi =j_z
\end{array}
\right|  \label{a1}
\end{equation}

has got the following type:

\[
\left\{ 
\begin{array}{c}
\psi ^{\dagger }\cdot \psi =\rho \mbox{,} \\ 
\left| \psi \right| ^2\cdot \left( 
\begin{array}{c}
\cos ^2\left( a\right) \cdot \sin \left( 2\cdot b\right) \cdot \cos \left(
d-g\right) - \\ 
-\sin ^2\left( a\right) \cdot \sin \left( 2\cdot v\right) \cdot \cos \left(
q-f\right)
\end{array}
\right) =j_x\mbox{,} \\ 
\left| \psi \right| ^2\cdot \left( 
\begin{array}{c}
\cos ^2\left( a\right) \cdot \sin \left( 2\cdot b\right) \cdot \sin \left(
d-g\right) - \\ 
-\sin ^2\left( a\right) \cdot \sin \left( 2\cdot v\right) \cdot \sin \left(
q-f\right)
\end{array}
\right) =j_y\mbox{,} \\ 
\left| \psi \right| ^2\cdot \left( \cos ^2\left( a\right) \cdot \cos \left(
2\cdot b\right) -\sin ^2\left( a\right) \cdot \cos \left( 2\cdot v\right)
\right) =j_z
\end{array}
\right| \mbox{.} 
\]

Hence for every probability current vector $\left\langle \rho
,j_x,j_y,j_z\right\rangle $: the spinor $\psi $, obeyed to this system,
exists.

If

\[
\mathbf{j}=\rho \cdot \mathbf{u} 
\]

then $\mathbf{u}$ is the average velocity.

The light pentad contains $\beta _4$, $\gamma _0$ besides $\beta _1$, $\beta
_2$, $\beta _3$ and yet two components of velocity can be defined as the
following:

\[
j_\xi =\psi ^{\dagger }\cdot \gamma _0\cdot \psi \mbox{, }j_\kappa =\psi
^{\dagger }\cdot \beta _4\cdot \psi \mbox{, }j_\xi =\rho \cdot u_\xi \cdot
\psi \mbox{, }j_\kappa =\rho \cdot u_\kappa . 
\]

In this case:

\[
\begin{array}{c}
u_\xi =\sin \left( 2\cdot a\right) \cdot \left[ \cos \left( b\right) \cdot
\cos \left( v\right) \cdot \cos \left( g-f\right) +\sin \left( b\right)
\cdot \sin \left( v\right) \cdot \cos \left( d-q\right) \right] \mbox{,} \\ 
u_\kappa =\sin \left( 2\cdot a\right) \cdot \left[ \cos \left( b\right)
\cdot \cos \left( v\right) \cdot \sin \left( g-f\right) +\sin \left(
b\right) \cdot \sin \left( v\right) \cdot \sin \left( d-q\right) \right] ;
\end{array}
\]

and:

\[
u_x^2+u_y^2+u_z^2+u_\xi ^2+u_\kappa ^2=1. 
\]

Hence for the completeness, yet two ''space'' coordinates $\xi $ and $\kappa 
$ should be added to our three $x,y,z$.

\section{The Hamiltonian}

The operator $\widehat{U}\left( t,\triangle t\right) $, which acts in the
set of these spinors, is denoted as the evolution operator for the spinor $%
\psi \left( t,\mathbf{x}\right) $, if:

\[
\psi \left( t+\triangle t,\mathbf{x}\right) =\widehat{U}\left( t,\triangle
t\right) \psi \left( t,\mathbf{x}\right) \mbox{.} 
\]

$\widehat{U}\left( t,\triangle t\right) $ is a linear operator.

The set of the spinors, for which $\widehat{U}\left( t,\triangle t\right) $
is the evolution operator, is denoted as the operator $\widehat{U}\left(
t,\triangle t\right) $ space.

The operator space is the linear space.

Let for an infinitesimal $\triangle t$:

\[
\widehat{U}\left( t,\triangle t\right) =1+\triangle t\cdot i\cdot \widehat{H}%
\left( t\right) \mbox{.} 
\]

Hence for an elements of the operator $\widehat{U}\left( t,\triangle
t\right) $ space:

\[
i\cdot \widehat{H}=\partial _t\mbox{.} 
\]

Since the functions $\rho $, $j_x$, $j_y$, $j_z$ fulfill to the continuity
equation:

\[
\partial _t\rho +\partial _xj_x+\partial _yj_y+\partial _zj_z=0 
\]

then:

\[
\left( \left( \partial _t\psi ^{\dagger }\right) \cdot \beta _0+\left(
\partial _x\psi ^{\dagger }\right) \cdot \beta _1+\left( \partial _y\psi
^{\dagger }\right) \cdot \beta _2+\left( \partial _z\psi ^{\dagger }\right)
\cdot \beta _3\right) \cdot \psi = 
\]

\[
=-\psi ^{\dagger }\cdot \left( \left( \beta _0\cdot \partial _t+\beta
_1\cdot \partial _x+\beta _2\cdot \partial _y+\beta _3\cdot \partial
_z\right) \psi \right) \mbox{.} 
\]

Let:

\[
\widehat{Q}=\left( i\cdot \widehat{H}+\beta _1\cdot \partial _x+\beta
_2\cdot \partial _y+\beta _3\cdot \partial _z\right) \mbox{.} 
\]

Hence:

\[
\psi ^{\dagger }\cdot \widehat{Q}^{\dagger }\cdot \psi =-\psi ^{\dagger
}\cdot \widehat{Q}\cdot \psi \mbox{.} 
\]

Therefore $i\cdot \widehat{Q}$ is the Hermitean operator.

Therefore:

\[
\widehat{H}=\beta _1\cdot \left( i\cdot \partial _x\right) +\beta _2\cdot
\left( i\cdot \partial _y\right) +\beta _3\cdot \left( i\cdot \partial
_z\right) -i\cdot \widehat{Q}\mbox{.} 
\]

Let

\[
-i\cdot \widehat{Q}= 
\]

\[
\left[ 
\begin{array}{cccc}
\varphi _{1,1} & \varphi _{1,2}+i\cdot \varpi _{1,2} & \varphi _{1,3}+i\cdot
\varpi _{1,3} & \varphi _{1,4}+i\cdot \varpi _{1,4} \\ 
\varphi _{1,2}-i\cdot \varpi _{1,2} & \varphi _{2,2} & \varphi _{2,3}+i\cdot
\varpi _{2,3} & \varphi _{2,4}+i\cdot \varpi _{2,4} \\ 
\varphi _{1,3}-i\cdot \varpi _{1,3} & \varphi _{2,3}-i\cdot \varpi _{2,3} & 
\varphi _{3,3} & \varphi _{3,4}+i\cdot \varpi _{3,4} \\ 
\varphi _{1,4}-i\cdot \varpi _{1,4} & \varphi _{2,4}-i\cdot \varpi _{2,4} & 
\varphi _{3,4}-i\cdot \varpi _{3,4} & \varphi _{4,4}
\end{array}
\right] \mbox{,} 
\]

here all $\varphi _{i,j}$ and $\varpi _{i,j}$ are a real functions on $%
R^{2+3+1}$.

In this paper I consider the case with:

\[
\begin{array}{c}
\varphi _{1,4}+i\cdot \varpi _{1,4}=0\mbox{,} \\ 
\varphi _{2,3}+i\cdot \varpi _{2,3}=0\mbox{,} \\ 
\varphi _{1,3}+i\cdot \varpi _{1,3}=\varphi _{2,4}+i\cdot \varpi _{2,4}%
\mbox{,}
\end{array}
\]

that is the electroweak part of SM \cite{GQ}, only.

Let $G_t$, $G_z$, $K_t$ and $K_z$ are the solution of the following system
of equations:

\[
\left\{ 
\begin{array}{c}
G_t+G_z+K_t+K_z=\varphi _{1,1}\mbox{,} \\ 
G_t-G_z+K_t-K_z=\varphi _{2,2}\mbox{,} \\ 
G_t-G_z-K_t+K_z=\varphi _{3,3}\mbox{,} \\ 
G_t+G_z-K_t-K_z=\varphi _{4,4};
\end{array}
\right| 
\]

$G_x$ and $K_x$ are the solution of the following system of equations:

\[
\left\{ 
\begin{array}{c}
G_x+K_x=\varphi _{1,2}\mbox{,} \\ 
-G_x+K_x=\varphi _{3,4}\mbox{;}
\end{array}
\right| 
\]

$G_y$ and $K_y$ are the solution of the following system of equations:

\[
\left\{ 
\begin{array}{c}
-G_y-K_y=\varpi _{1,2}\mbox{,} \\ 
G_y-K_x=\varpi _{3,4}\mbox{.}
\end{array}
\right| 
\]

In this case:

\[
\begin{array}{c}
-i\cdot \widehat{Q}= \\ 
=\left( G_t\cdot \beta _0+G_x\cdot \beta _1+G_y\cdot \beta _2+G_z\cdot \beta
_3\right) + \\ 
+\left( K_t\cdot \beta _0+K_x\cdot \beta _1+K_y\cdot \beta _2+K_z\cdot \beta
_3\right) \cdot \gamma _5+ \\ 
+\varphi _{1,3}\cdot \gamma _0+\varpi _{1,3}\cdot \beta _4\mbox{.}
\end{array}
\]

Therefore:

\[
\begin{array}{c}
\beta _0\cdot \left( \widehat{H}-G_t-K_t\cdot \gamma _5\right) \cdot \psi =
\\ 
=\beta _1\cdot i\cdot \left( \partial _x-iG_x-iK_x\cdot \gamma _5\right)
\cdot \psi + \\ 
+\beta _2\cdot i\cdot \left( \partial _y-iG_y-iK_y\cdot \gamma _5\right)
\cdot \psi + \\ 
+\beta _3\cdot i\cdot \left( \partial _z-iG_z-iK_z\cdot \gamma _5\right)
\cdot \psi + \\ 
+\gamma _0\cdot \varphi _{1,3}\cdot \psi +\beta _4\cdot \varpi _{1,3}\cdot
\psi \mbox{.}
\end{array}
\]

Let $R_\xi $ and $R_\kappa $ be a functions for which:

\begin{eqnarray*}
R_\xi \psi &=&\varphi _{1,3}\psi -i\partial _\xi \psi \mbox{,} \\
R_\kappa \psi &=&\varpi _{1,3}\psi -i\partial _\kappa \psi \mbox{.}
\end{eqnarray*}

In this case (similar to \cite{RLI}):

\begin{equation}
\begin{array}{c}
\beta _0\cdot \left( \widehat{H}-G_t-K_t\cdot \gamma _5\right) = \\ 
=\beta _1\cdot i\cdot \left( \partial _x-iG_x-iK_x\cdot \gamma _5\right) +
\\ 
+\beta _2\cdot i\cdot \left( \partial _y-iG_y-iK_y\cdot \gamma _5\right) +
\\ 
+\beta _3\cdot i\cdot \left( \partial _z-iG_z-iK_z\cdot \gamma _5\right) +
\\ 
+\gamma _0\cdot i\cdot \left( \partial _\xi -iR_\xi \right) + \\ 
+\beta _4\cdot i\cdot \left( \partial _\kappa -iR_\kappa \right) \mbox{.}
\end{array}
\label{e1}
\end{equation}

The equation of motion (the Euler-Lagrange equation) is:

\begin{equation}
\begin{array}{c}
\beta _0\cdot i\partial _t\psi +\beta _1\cdot i\partial _x\psi +\beta
_2\cdot i\partial _y\psi +\beta _3\cdot i\partial _z\psi + \\ 
+\beta _5\cdot i\partial _\xi \psi +\beta _4\cdot i\partial _\kappa \psi +
\\ 
+\beta _0\cdot G_t\cdot \psi +\beta _1\cdot G_x\cdot \psi +\beta _2\cdot
G_y\cdot \psi +\beta _3\cdot G_z\cdot \psi + \\ 
+\beta _5\cdot R_\xi \cdot \psi +\beta _4\cdot R_\kappa \cdot \psi + \\ 
+K_t\cdot \gamma _5\cdot \psi + \\ 
+\beta _1\cdot K_x\cdot \gamma _5\cdot \psi +\beta _2\cdot K_y\cdot \gamma
_5\cdot \psi +\beta _3\cdot K_z\cdot \gamma _5\cdot \psi = \\ 
=0\mbox{.}
\end{array}
\label{a5}
\end{equation}

\section{The Global Rotations}

\subsection{The rotation of $yOz$, $xOz$ and $xOy$}

Let $\alpha $ be a constant real number and:

\[
\begin{array}{c}
y^{\prime }=y\cdot \cos 2\alpha +z\cdot \sin 2\alpha \mbox{,} \\ 
z^{\prime }=z\cdot \cos 2\alpha -y\cdot \sin 2\alpha \mbox{,} \\ 
t^{\prime }=t, \\ 
x^{\prime }=y, \\ 
\xi ^{\prime }=\xi , \\ 
\kappa ^{\prime }=\kappa \mbox{.}
\end{array}
\]

In this case:

\begin{equation}
\begin{array}{c}
j_{y^{\prime }}^{\prime }=j_y\cdot \cos 2\alpha +j_z\cdot \sin 2\alpha %
\mbox{,} \\ 
j_{z^{\prime }}^{\prime }=j_z\cdot \cos 2\alpha -j_y\cdot \sin 2\alpha %
\mbox{,} \\ 
\rho ^{\prime }=\rho \mbox{,} \\ 
j_{x^{\prime }}^{\prime }=j_x\mbox{,} \\ 
j_{\xi ^{\prime }}^{\prime }=j_\xi \mbox{,} \\ 
j_{\kappa ^{\prime }}^{\prime }=j_\kappa \mbox{.}
\end{array}
\label{a2}
\end{equation}

Let:

\[
U_{2,3}=\cos \alpha \cdot 1_4+\sin \alpha \cdot \beta _2\cdot \beta _3 
\]

and

\begin{equation}
\psi ^{\prime }=U_{2,3}\psi \mbox{.}  \label{a4}
\end{equation}

Because:

\[
\begin{array}{c}
U_{2,3}^{\dagger }\beta _2U_{2,3}=\beta _2\cdot \cos 2\alpha +\beta _3\cdot
\sin 2\alpha \mbox{,} \\ 
U_{2,3}^{\dagger }\beta _3U_{2,3}=\beta _3\cdot \cos 2\alpha -\beta _2\cdot
\sin 2\alpha \mbox{,} \\ 
U_{2,3}^{\dagger }U_{2,3}=1_4\mbox{,} \\ 
U_{2,3}^{\dagger }\beta _1U_{2,3}=\beta _1\mbox{,} \\ 
U_{2,3}^{\dagger }\beta _4U_{2,3}=\beta _4\mbox{,} \\ 
U_{2,3}^{\dagger }\beta _5U_{2,3}=\beta _5\mbox{,}
\end{array}
\]

then from (\ref{a1}) and (\ref{a2}):

\[
\left\{ 
\begin{array}{c}
\psi ^{\prime \dagger }\cdot \psi ^{\prime }=\rho ^{\prime }\mbox{,} \\ 
\psi ^{\prime \dagger }\cdot \beta _1\cdot \psi ^{\prime }=j_{x^{\prime
}}^{\prime }\mbox{,} \\ 
\psi ^{\prime \dagger }\cdot \beta _2\cdot \psi ^{\prime }=j_{y^{\prime
}}^{\prime }\mbox{,} \\ 
\psi ^{\prime \dagger }\cdot \beta _3\cdot \psi ^{\prime }=j_{z^{\prime
}}^{\prime }
\end{array}
\right| 
\]

and the equation of motion is:

\begin{equation}
\begin{array}{c}
\beta _0\cdot i\partial _{t^{\prime }}\psi ^{\prime }+\beta _1\cdot
i\partial _{x^{\prime }}\psi ^{\prime }+\beta _2\cdot i\partial _{y^{\prime
}}\psi ^{\prime }+\beta _3\cdot i\partial _{z^{\prime }}\psi ^{\prime }+ \\ 
+\beta _5\cdot i\partial _{\xi ^{\prime }}\psi ^{\prime }+\beta _4\cdot
i\partial _{\kappa ^{\prime }}\psi ^{\prime }+ \\ 
+\beta _0\cdot G_{t^{\prime }}^{\prime }\cdot \psi ^{\prime }+\beta _1\cdot
G_{x^{\prime }}^{\prime }\cdot \psi ^{\prime }+\beta _2\cdot G_{y^{\prime
}}^{\prime }\cdot \psi ^{\prime }+\beta _3\cdot G_{z^{\prime }}^{\prime
}\cdot \psi ^{\prime }+ \\ 
+\beta _5\cdot R_{\xi ^{\prime }}^{\prime }\cdot \psi ^{\prime }+\beta
_4\cdot R_{\kappa ^{\prime }}^{\prime }\cdot \psi ^{\prime }+ \\ 
+K_{t^{\prime }}^{\prime }\cdot \gamma _5\cdot \psi ^{\prime }+ \\ 
+\beta _1\cdot K_{x^{\prime }}^{\prime }\cdot \gamma _5\cdot \psi ^{\prime
}+\beta _2\cdot K_{y^{\prime }}^{\prime }\cdot \gamma _5\cdot \psi ^{\prime
}+\beta _3\cdot K_{z^{\prime }}^{\prime }\cdot \gamma _5\cdot \psi ^{\prime
}= \\ 
=0\mbox{.}
\end{array}
\label{a3}
\end{equation}

Since:

\begin{eqnarray*}
\beta _2\cdot U_{2,3} &=&U_{2,3}\cdot \left( \beta _2\cdot \cos 2\alpha
+\beta _3\cdot \sin 2\alpha \right) \mbox{,} \\
\beta _3\cdot U_{2,3} &=&U_{2,3}\cdot \left( \beta _3\cdot \cos 2\alpha
-\beta _2\cdot \sin 2\alpha \right) \mbox{,} \\
\beta _1\cdot U_{2,3} &=&U_{2,3}\cdot \beta _1\mbox{,} \\
\beta _4\cdot U_{2,3} &=&U_{2,3}\cdot \beta _4\mbox{,} \\
\beta _5\cdot U_{2,3} &=&U_{2,3}\cdot \beta _5
\end{eqnarray*}

and

\[
\begin{array}{c}
\partial _{y^{\prime }}\psi =\partial _y\psi \cdot \cos 2\alpha +\partial
_z\psi \cdot \sin 2\alpha \mbox{,} \\ 
\partial _{z^{\prime }}\psi =-\partial _y\psi \cdot \sin 2\alpha +\partial
_z\psi \cdot \cos 2\alpha
\end{array}
\]

then from (\ref{a3}) by (\ref{a4}) the following equation is derived:

\[
\begin{array}{c}
\beta _0\cdot i\partial _t\psi +\beta _1\cdot i\partial _x\psi +\beta
_2\cdot i\partial _y\psi +\beta _3\cdot i\partial _z\psi +\beta _5\cdot
i\partial _\xi \psi +\beta _4\cdot i\partial _\kappa \psi \\ 
+\beta _0\cdot G_{t^{\prime }}^{\prime }\cdot \psi +\beta _1\cdot
G_{x^{\prime }}^{\prime }\cdot \psi + \\ 
+\beta _2\cdot \left( G_{y^{\prime }}^{\prime }\cdot \cos 2\alpha
-G_{z^{\prime }}^{\prime }\cdot \sin 2\alpha \right) \cdot \psi + \\ 
+\beta _3\cdot \left( G_{z^{\prime }}^{\prime }\cdot \cos 2\alpha
+G_{y^{\prime }}^{\prime }\cdot \sin 2\alpha \right) \cdot \psi + \\ 
+K_{t^{\prime }}^{\prime }\cdot \gamma _5\cdot \psi +\beta _1\cdot
K_{x^{\prime }}^{\prime }\cdot \gamma _5\cdot \psi + \\ 
+\beta _2\cdot \left( K_{y^{\prime }}^{\prime }\cdot \cos 2\alpha
-K_{z^{\prime }}^{\prime }\cdot \sin 2\alpha \right) \cdot \gamma _5\cdot
\psi + \\ 
+\beta _3\cdot \left( K_{z^{\prime }}^{\prime }\cdot \cos 2\alpha
+K_{y^{\prime }}^{\prime }\cdot \sin 2\alpha \right) \cdot \gamma _5\cdot
\psi + \\ 
+\beta _5\cdot R_{\xi ^{\prime }}^{\prime }\cdot \psi +\beta _4\cdot
R_{\kappa ^{\prime }}^{\prime }\cdot \psi =0\mbox{.}
\end{array}
\]

From it and from (\ref{a5}):

\[
\begin{array}{c}
G_y=G_{y^{\prime }}^{\prime }\cdot \cos 2\alpha -G_{z^{\prime }}^{\prime
}\cdot \sin 2\alpha \mbox{,} \\ 
G_z=G_{z^{\prime }}^{\prime }\cdot \cos 2\alpha +G_{y^{\prime }}^{\prime
}\cdot \sin 2\alpha \mbox{,} \\ 
G_t=G_{t^{\prime }}^{\prime }\mbox{,} \\ 
G_x=G_{x^{\prime }}^{\prime }\mbox{,} \\ 
R_\xi =R_{\xi ^{\prime }}^{\prime }\mbox{,} \\ 
R_{\kappa ^{\prime }}^{\prime }=R_\kappa
\end{array}
\]

and

\[
\begin{array}{c}
K_y=K_{y^{\prime }}^{\prime }\cdot \cos 2\alpha -K_{z^{\prime }}^{\prime
}\cdot \sin 2\alpha \mbox{,} \\ 
K_z=K_{z^{\prime }}^{\prime }\cdot \cos 2\alpha +K_{y^{\prime }}^{\prime
}\cdot \sin 2\alpha \mbox{,} \\ 
K_t=K_{t^{\prime }}^{\prime }\mbox{,} \\ 
K_x=K_{x^{\prime }}^{\prime }\mbox{.}
\end{array}
\]

Like to above, the rotation of $xOz$ and $xOy$ behave with:

\[
U_{1,3}=\cos \alpha \cdot 1_4+\sin \alpha \cdot \beta _1\cdot \beta _3 
\]

and

\[
U_{1,2}=\cos \alpha \cdot 1_4+\sin \alpha \cdot \beta _1\cdot \beta _2 
\]

with accordance.

Therefore the triplets $\left\{ G_x,G_y,G_z\right\} $ and $\left\{
K_x,K_y,K_z\right\} $ behave as \\a 3-vectors for the rotations of the
Cartesian coordinates system.

\subsection{The rotation of $tOx$, $tOy$ and $tOz$}

Let $v$ be a real constant number, for which: $\left| v\right| <1$.

Let $\alpha $ be a real number, for which:

\[
\sinh \left( \alpha \right) =\frac v{\sqrt{1-v^2}}\mbox{.} 
\]

Let:

\begin{eqnarray*}
t &=&t"\cdot \cosh 2\alpha -x"\cdot \sinh 2\alpha \mbox{,} \\
x &=&x"\cdot \cosh 2\alpha -t"\cdot \sinh 2\alpha \mbox{,} \\
y &=&y"\mbox{,} \\
z &=&z"\mbox{,} \\
\xi &=&\xi "\mbox{,} \\
\kappa &=&\kappa "\mbox{.}
\end{eqnarray*}

In this case:

\begin{equation}
\begin{array}{c}
\rho "=\rho \cdot \cosh 2\alpha +j_x\cdot \sinh 2\alpha \mbox{,} \\ 
j_{x"}"=j_x\cdot \cosh 2\alpha +\rho \cdot \sinh 2\alpha \mbox{,} \\ 
j_{y"}"=j_y\mbox{,} \\ 
j_{z"}"=j_z\mbox{,} \\ 
j_{\xi "}"=j_\xi \mbox{,} \\ 
j_{\kappa "}"=j_\kappa \mbox{.}
\end{array}
\label{b1}
\end{equation}

Let

\[
U_{0,1}=\cosh \alpha \cdot \beta _0+\sinh \alpha \cdot \beta _0\cdot \beta
_1 
\]

and

\[
\psi "=U_{0,1}\psi \mbox{.} 
\]

Because

\begin{equation}
\begin{array}{c}
U_{0,1}^{\dagger }\cdot \beta _0\cdot U_{0,1}=\cosh 2\alpha \cdot \beta
_0+\sinh 2\alpha \cdot \beta _1 \\ 
U_{0,1}^{\dagger }\cdot \beta _1\cdot U_{0,1}=\cosh 2\alpha \cdot \beta
_1+\sinh 2\alpha \cdot \beta _0 \\ 
U_{0,1}^{\dagger }\cdot \beta _2\cdot U_{0,1}=\beta _2 \\ 
U_{0,1}^{\dagger }\cdot \beta _3\cdot U_{0,1}=\beta _3 \\ 
U_{0,1}^{\dagger }\cdot \beta _5\cdot U_{0,1}=\beta _5 \\ 
U_{0,1}^{\dagger }\cdot \beta _4\cdot U_{0,1}=\beta _4
\end{array}
\label{b2}
\end{equation}

then from (\ref{a1}) and (\ref{b1}):

\[
\left\{ 
\begin{array}{c}
\psi "\cdot \psi "=\rho "\mbox{,} \\ 
\psi "\cdot \beta _1\cdot \psi "=j_{x"}"\mbox{,} \\ 
\psi "\cdot \beta _2\cdot \psi "=j_{y"}"\mbox{,} \\ 
\psi "\cdot \beta _3\cdot \psi "=j_{z"}"
\end{array}
\right| 
\]

and the equation of motion is:

\[
\begin{array}{c}
\beta _0\cdot i\partial _{t"}\left( U_{0,1}\psi \right) +\beta _1\cdot
i\partial _{x"}\left( U_{0,1}\psi \right) + \\ 
+\beta _2\cdot i\partial _{y"}\left( U_{0,1}\psi \right) +\beta _3\cdot
i\partial _{z"}\left( U_{0,1}\psi \right) + \\ 
+\beta _5\cdot i\partial _{\xi "}\left( U_{0,1}\psi \right) +\beta _4\cdot
i\partial _{\kappa "}\left( U_{0,1}\psi \right) + \\ 
+\beta _0\cdot G_{t"}"\cdot \left( U_{0,1}\psi \right) +\beta _1\cdot
G_{x"}"\cdot \left( U_{0,1}\psi \right) + \\ 
+\beta _2\cdot G_{y"}"\cdot \left( U_{0,1}\psi \right) +\beta _3\cdot
G_{z"}"\cdot \left( U_{0,1}\psi \right) + \\ 
+\beta _5\cdot R_{\xi "}"\cdot \left( U_{0,1}\psi \right) +\beta _4\cdot
R_{\kappa "}"\cdot \left( U_{0,1}\psi \right) + \\ 
+K_{t"}"\cdot \gamma _5\cdot \left( U_{0,1}\psi \right) +\beta _1\cdot
K_{x"}"\cdot \gamma _5\cdot \left( U_{0,1}\psi \right) + \\ 
+\beta _2\cdot K_{y"}"\cdot \gamma _5\cdot \left( U_{0,1}\psi \right) +\beta
_3\cdot K_{z"}"\cdot \gamma _5\cdot \left( U_{0,1}\psi \right) = \\ 
=0\mbox{;}
\end{array}
\]

hence:

\[
\begin{array}{c}
U_{0,1}^{\dagger }\cdot U_{0,1}\cdot i\partial _{t"}\psi +U_{0,1}^{\dagger
}\cdot \beta _1\cdot U_{0,1}\cdot i\partial _{x"}\psi + \\ 
+U_{0,1}^{\dagger }\cdot \beta _2\cdot U_{0,1}\cdot i\partial _{y"}\psi
+U_{0,1}^{\dagger }\cdot \beta _3\cdot U_{0,1}\cdot i\partial _{z"}\psi + \\ 
+U_{0,1}^{\dagger }\cdot \beta _5\cdot U_{0,1}\cdot i\partial _{\xi "}\psi
+U_{0,1}^{\dagger }\cdot \beta _4\cdot U_{0,1}\cdot i\partial _{\kappa
"}\psi + \\ 
+U_{0,1}^{\dagger }\cdot U_{0,1}\cdot G_{t"}"\cdot \psi +U_{0,1}^{\dagger
}\cdot \beta _1\cdot U_{0,1}\cdot G_{x"}"\cdot \psi + \\ 
+U_{0,1}^{\dagger }\cdot \beta _2\cdot U_{0,1}\cdot G_{y"}"\cdot \psi
+U_{0,1}^{\dagger }\cdot \beta _3\cdot U_{0,1}\cdot G_{z"}"\cdot \psi + \\ 
+U_{0,1}^{\dagger }\cdot \beta _5\cdot U_{0,1}\cdot R_{\xi "}"\cdot \psi
+U_{0,1}^{\dagger }\cdot \beta _4\cdot U_{0,1}\cdot R_{\kappa "}"\cdot \psi +
\\ 
+U_{0,1}^{\dagger }\cdot K_{t"}"\cdot \gamma _5\cdot U_{0,1}\cdot \psi + \\ 
+U_{0,1}^{\dagger }\cdot \beta _1\cdot K_{x"}"\cdot \gamma _5\cdot
U_{0,1}\cdot \psi + \\ 
+U_{0,1}^{\dagger }\cdot \beta _2\cdot K_{y"}"\cdot \gamma _5\cdot
U_{0,1}\cdot \psi + \\ 
+U_{0,1}^{\dagger }\cdot \beta _3\cdot K_{z"}"\cdot \gamma _5\cdot
U_{0,1}\cdot \psi = \\ 
=0\mbox{.}
\end{array}
\]

Since

\[
\gamma _5\cdot U_{0,1}=U_{0,1}\cdot \gamma _5 
\]

and

\begin{eqnarray*}
\partial _{t"}\psi &=&\left( \partial _t\psi \cdot \cosh 2\alpha -\partial
_x\psi \cdot \sinh 2\alpha \right) \mbox{,} \\
\partial _{x"}\psi &=&\left( \partial _x\psi \cdot \cosh 2\alpha -\partial
_t\psi \cdot \sinh 2\alpha \right)
\end{eqnarray*}

then from (\ref{b2})

\[
\begin{array}{c}
\beta _0\cdot i\partial _t\psi +\beta _1\cdot i\partial _x\psi +\beta
_2\cdot i\partial _y\psi +\beta _3\cdot i\partial _z\psi +\beta _5\cdot
i\partial _\xi \psi +\beta _4\cdot i\partial _\kappa \psi + \\ 
+\beta _0\cdot \left( \cosh 2\alpha \cdot G_{t"}"+\sinh 2\alpha \cdot
G_{x"}"\right) \cdot \psi + \\ 
+\beta _1\cdot \left( \cosh 2\alpha \cdot G_{x"}"+\sinh 2\alpha \cdot
G_{t"}"\right) \cdot \psi + \\ 
+\beta _2\cdot G_{y"}"\cdot \psi +\beta _3\cdot G_{z"}"\cdot \psi +\beta
_5\cdot R_{\xi "}"\cdot \psi +\beta _4\cdot R_{\kappa "}"\cdot \psi + \\ 
+\left( \cosh 2\alpha \cdot K_{t"}"+\sinh 2\alpha \cdot K_{x"}"\right) \cdot
\gamma _5\cdot \psi + \\ 
+\beta _1\cdot \left( \cosh 2\alpha \cdot K_{x"}"+\sinh 2\alpha \cdot
K_{t"}"\right) \cdot \gamma _5\cdot \psi + \\ 
+\beta _2\cdot K_{y"}"\cdot \gamma _5\cdot \psi +\beta _3\cdot K_{z"}"\cdot
\gamma _5\cdot \psi = \\ 
=0\mbox{.}
\end{array}
\]

From it and from (\ref{a5}):

\[
\begin{array}{c}
G_t=\cosh 2\alpha \cdot G_{t"}"+\sinh 2\alpha \cdot G_{x"}"\mbox{,} \\ 
G_x=\cosh 2\alpha \cdot G_{x"}"+\sinh 2\alpha \cdot G_{t"}"\mbox{,} \\ 
G_y=G_{y"}"\mbox{,} \\ 
G_z=G_{z"}"\mbox{,} \\ 
R_\xi =R_{\xi "}"\mbox{,} \\ 
R_\kappa =R_{\kappa "}"
\end{array}
\]

and

\[
\begin{array}{c}
K_t=\cosh 2\alpha \cdot K_{t"}"+\sinh 2\alpha \cdot K_{x"}"\mbox{,} \\ 
K_x=\cosh 2\alpha \cdot K_{x"}"+\sinh 2\alpha \cdot K_{t"}"\mbox{,} \\ 
K_y=K_{y"}"\mbox{,} \\ 
K_z=K_{z"}"\mbox{.}
\end{array}
\]

Like to above, the rotation of $tOz$ and $tOy$ behave with:

\[
U_{0,3}=\cosh \alpha \cdot \beta _0+\sinh \alpha \cdot \beta _0\cdot \beta
_3 
\]

and

\[
U_{0,2}=\cosh \alpha \cdot \beta _0+\sinh \alpha \cdot \beta _0\cdot \beta
_2 
\]

with accordance.

\section{The Local Rotations of $\xi O\kappa $}

Let $\theta $ be a real function of $\left( t,x,y,z,\xi ,\kappa \right) $
for which: $\partial _\xi \theta =0$ and $\partial _\kappa \theta =0$ and:

\[
\begin{array}{c}
t`=t\mbox{,} \\ 
x`=x\mbox{,} \\ 
y`=y\mbox{,} \\ 
z`=z\mbox{,} \\ 
\xi `=\xi \cdot \cos 2\theta +\kappa \cdot \sin 2\theta \mbox{,} \\ 
\kappa `=\kappa \cdot \cos 2\theta -\xi \cdot \sin 2\theta \mbox{.}
\end{array}
\]

(This transformation corresponds to the weak isospin transformation \cite{PV}%
)

In this case:

\[
\begin{array}{c}
\rho `=\rho \mbox{,} \\ 
j_x`=j_x\mbox{,} \\ 
j_y`=j_y\mbox{,} \\ 
j_z`=j_z\mbox{,} \\ 
j_\xi `=j_\xi \cdot \cos 2\theta +j_\kappa \cdot \sin 2\theta \mbox{,} \\ 
j_\kappa `=j_\kappa \cdot \cos 2\theta -j_\xi \cdot \sin 2\theta \mbox{.}
\end{array}
\]

Let:

\[
U_{5,4}=\cos \theta \cdot 1_4+\sin \theta \cdot \beta _5\cdot \beta _4 
\]

and

\[
\psi `=\left( U_{5,4}\psi \right) \mbox{.} 
\]

Because

\begin{equation}
\begin{array}{c}
U_{5,4}^{\dagger }\cdot U_{5,4}=1_4\mbox{,} \\ 
U_{5,4}^{\dagger }\cdot \beta _1\cdot U_{5,4}=\beta _1\mbox{,} \\ 
U_{5,4}^{\dagger }\cdot \beta _2\cdot U_{5,4}=\beta _2\mbox{,} \\ 
U_{5,4}^{\dagger }\cdot \beta _3\cdot U_{5,4}=\beta _3\mbox{,} \\ 
U_{5,4}^{\dagger }\cdot \beta _5\cdot U_{5,4}=\beta _5\cdot \cos 2\theta
+\beta _4\cdot \sin 2\theta \mbox{,} \\ 
U_{5,4}^{\dagger }\cdot \beta _4\cdot U_{5,4}=\beta _4\cdot \cos 2\theta
-\beta _5\cdot \sin 2\theta
\end{array}
\label{c1}
\end{equation}

then from above and from (\ref{a1}):

\[
\begin{array}{c}
\psi `^{\dagger }\cdot \psi `=\rho `\mbox{,} \\ 
\psi `^{\dagger }\cdot \beta _1\cdot \psi `=j_x`\mbox{,} \\ 
\psi `^{\dagger }\cdot \beta _2\cdot \psi `=j_y`\mbox{,} \\ 
\psi `^{\dagger }\cdot \beta _3\cdot \psi `=j_z`\mbox{.}
\end{array}
\]

and the equation of motion is:

\[
\begin{array}{c}
\beta _0\cdot i\partial _{t`}\left( U_{5,4}\psi \right) +\beta _1\cdot
i\partial _{x`}\left( U_{5,4}\psi \right) +\beta _2\cdot i\partial
_{y`}\left( U_{5,4}\psi \right) +\beta _3\cdot i\partial _{z`}\left(
U_{5,4}\psi \right) + \\ 
+\beta _5\cdot i\partial _{\xi `}\left( U_{5,4}\psi \right) +\beta _4\cdot
i\partial _{\kappa `}\left( U_{5,4}\psi \right) + \\ 
+\beta _0\cdot G_{t`}`\cdot \left( U_{5,4}\psi \right) ++\beta _1\cdot
G_{x`}`\cdot \left( U_{5,4}\psi \right) + \\ 
+\beta _2\cdot G_{y`}`\cdot \left( U_{5,4}\psi \right) +\beta _3\cdot
G_{z`}`\cdot \left( U_{5,4}\psi \right) + \\ 
+\beta _5\cdot R_{\xi `}`\cdot \left( U_{5,4}\psi \right) +\beta _4\cdot
R_{\kappa `}`\cdot \left( U_{5,4}\psi \right) + \\ 
+K_{t`}`\cdot \gamma _5\cdot \left( U_{5,4}\psi \right) +\beta _1\cdot
K_{x`}`\cdot \gamma _5\cdot \left( U_{5,4}\psi \right) + \\ 
+\beta _2\cdot K_{y`}`\cdot \gamma _5\cdot \left( U_{5,4}\psi \right) +\beta
_3\cdot K_{z`}`\cdot \gamma _5\cdot \left( U_{5,4}\psi \right) = \\ 
=0\mbox{.}
\end{array}
\]

Since

\[
\gamma _5U_{5,4}=U_{5,4}\gamma _5 
\]

and 
\[
\partial _\mu U_{5,4}=-i\partial _\mu \theta \cdot \gamma _5\cdot U_{5,4} 
\]

for all variables $\mu $ then

\[
\begin{array}{c}
U_{5,4}\cdot \partial _t\theta \cdot \gamma _5\cdot \psi +U_{5,4}\cdot \beta
_0\cdot i\partial _t\psi +\beta _1U_{5,4}\cdot \partial _x\theta \cdot
\gamma _5\cdot \psi +\beta _1U_{5,4}\cdot i\partial _x\psi + \\ 
+\beta _2U_{5,4}\cdot \partial _y\theta \cdot \gamma _5\cdot \psi +\beta
_2U_{5,4}\cdot i\partial _y\psi +\beta _3U_{5,4}\cdot \partial _z\theta
\cdot \gamma _5\cdot \psi +\beta _3U_{5,4}\cdot i\partial _z\psi + \\ 
+\beta _5U_{5,4}\cdot i\partial _{\xi `}\psi +\beta _4U_{5,4}\cdot i\partial
_{\kappa `}\psi + \\ 
+G_{t`}`\cdot U_{5,4}\psi +\beta _1U_{5,4}\cdot G_{x`}`\cdot \psi +\beta
_2U_{5,4}\cdot G_{y`}`\cdot \psi +\beta _3U_{5,4}\cdot G_{z`}`\cdot \psi +
\\ 
+\beta _5U_{5,4}\cdot R_{\xi `}`\cdot \psi +\beta _4U_{5,4}\cdot R_{\kappa
`}`\cdot \psi + \\ 
+U_{5,4}\cdot K_{t`}`\cdot \gamma _5\cdot \psi + \\ 
+\beta _1U_{5,4}\cdot K_{x`}`\cdot \gamma _5\cdot \psi +\beta _2U_{5,4}\cdot
K_{y`}`\cdot \gamma _5\cdot \psi +\beta _3U_{5,4}\cdot K_{z`}`\cdot \gamma
_5\cdot \psi = \\ 
=0\mbox{.}
\end{array}
\]

Since

\[
\begin{array}{c}
\partial _{\xi `}\psi =\partial _\xi \psi \cdot \cos 2\theta +\partial
_\kappa \psi \cdot \sin 2\theta \mbox{,} \\ 
\partial _{\kappa `}\psi =\partial _\kappa \psi \cdot \cos 2\theta -\partial
_\xi \psi \cdot \sin 2\theta
\end{array}
\]

then from above and from (\ref{c1}):

\[
\begin{array}{c}
\beta _0\cdot i\partial _t\psi +\beta _1\cdot i\partial _x\psi +\beta
_2\cdot i\partial _y\psi +\beta _3\cdot i\partial _z\psi + \\ 
+\beta _5\cdot i\partial _\xi \psi +\beta _4\cdot i\partial _\kappa \psi +
\\ 
+\beta _0\cdot G_{t`}`\cdot \psi +\beta _1\cdot G_{x`}`\cdot \psi +\beta
_2\cdot G_{y`}`\cdot \psi +\beta _3\cdot G_{z`}`\cdot \psi + \\ 
+\beta _5\cdot \left( \cos 2\theta \cdot R_{\xi `}`-\sin 2\theta \cdot
R_4`\right) \cdot \psi + \\ 
+\beta _4\cdot \left( \cos 2\theta \cdot R_{\kappa `}`+\sin 2\theta \cdot
R_{\xi `}`\right) \cdot \psi + \\ 
+\left( K_{t`}`+\partial _t\theta \right) \cdot \gamma _5\cdot \psi +\beta
_1\cdot \left( K_{x`}`+\partial _x\theta \right) \cdot \gamma _5\cdot \psi +
\\ 
+\beta _2\cdot \left( K_{y`}`+\partial _y\theta \right) \cdot \gamma _5\cdot
\psi +\beta _3\cdot \left( K_{z`}`+\partial _z\theta \right) \cdot \gamma
_5\cdot \psi = \\ 
=0\mbox{.}
\end{array}
\]

From above and from (\ref{a5}):

\[
\begin{array}{c}
K_{t`}`+\partial _t\theta =K_t\mbox{,} \\ 
K_{x`}`+\partial _x\theta =K_x\mbox{,} \\ 
K_{y`}`+\partial _y\theta =K_y\mbox{,} \\ 
K_{z`}`+\partial _z\theta =K_z\mbox{,} \\ 
\cos 2\theta \cdot R_{\xi `}`-\sin 2\theta \cdot R_{\kappa `}`=R_\xi \mbox{,}
\\ 
\cos 2\theta \cdot R_{\kappa `}`+\sin 2\theta \cdot R_{\xi `}`=R_\kappa %
\mbox{.}
\end{array}
\]

\section{The Lagrangians}

If

\[
K_\xi =Re\left( R_\xi \right) \mbox{ and }K_\kappa =Re\left( R_\kappa
\right) 
\]

then the Lagrangian for the Hamiltonian (\ref{e1}) is the following:

\[
\mathcal{L}_f=0.5\cdot i\cdot 
\]

\[
\cdot \left( 
\begin{array}{c}
\psi ^{\dagger }\beta _0\partial _t\psi +\psi ^{\dagger }\beta _1\partial
_x\psi +\psi ^{\dagger }\beta _2\partial _y\psi +\psi ^{\dagger }\beta
_3\partial _z\psi +\psi ^{\dagger }\beta _5\partial _\xi \psi +\psi
^{\dagger }\beta _4\partial _\kappa \psi - \\ 
-\left( \partial _t\psi \right) ^{\dagger }\beta _0\psi -\left( \partial
_x\psi \right) ^{\dagger }\beta _1\psi -\left( \partial _y\psi \right)
^{\dagger }\beta _2\psi -\left( \partial _z\psi \right) ^{\dagger }\beta
_3\psi - \\ 
-\left( \partial _\xi \psi \right) ^{\dagger }\beta _5\psi -\left( \partial
_\kappa \psi \right) ^{\dagger }\beta _4\psi
\end{array}
\right) + 
\]

\[
+\left( 
\begin{array}{c}
\psi ^{\dagger }\beta _0G_t\psi +\psi ^{\dagger }\beta _1G_x\psi +\psi
^{\dagger }\beta _2G_y\psi +\psi ^{\dagger }\beta _3G_z\psi + \\ 
+\psi ^{\dagger }K_t\gamma _5\psi +\psi ^{\dagger }\beta _1K_x\gamma _5\psi
+\psi ^{\dagger }\beta _2K_y\gamma _5\psi +\psi ^{\dagger }\beta _3K_z\gamma
_5\psi + \\ 
+\psi ^{\dagger }\beta _5K_\xi \psi +\psi ^{\dagger }\beta _4K_\kappa \psi
\end{array}
\right) \mbox{.} 
\]

If

\[
H_{\mu ,\nu }=\partial _\mu K_\nu -\partial _\nu K_\mu 
\]

and

\[
H= 
\]

\[
\left[ 
\begin{array}{cccccc}
0 & H_{t,x} & H_{t,y} & H_{t,z} & H_{t,\kappa } & H_{t,\xi } \\ 
-H_{t,x} & 0 & H_{x,y} & H_{x,z} & H_{x,\kappa } & H_{x,\xi } \\ 
-H_{t,y} & -H_{x,y} & 0 & H_{y,z} & H_{y,\kappa } & H_{y,\xi } \\ 
-H_{t,z} & -H_{x,z} & -H_{y,z} & 0 & H_{z,\kappa } & H_{z,\xi } \\ 
-H_{t,\kappa } & -H_{x,\kappa } & -H_{y,\kappa } & -H_{z,\kappa } & 0 & 
H_{\kappa ,\xi } \\ 
-H_{t,\xi } & -H_{x,\xi } & -H_{y,\xi } & -H_{z,\xi } & -H_{\kappa ,\xi } & 0
\end{array}
\right] 
\]

then the Lagrangian for $K_\mu $ ($0\leq \mu \leq 5$):

\[
\mathcal{L}_K=H^{\dagger }\cdot H\mbox{.} 
\]

The Lagrange-Euler equation for $K_\mu $ is:

\[
\begin{array}{c}
\partial _t\partial _tK_\mu +\partial _x\partial _xK_\mu +\partial
_y\partial _yK_\mu +\partial _z\partial _zK_\mu +\partial _\xi \partial _\xi
K_\mu +\partial _\kappa \partial _\kappa K_\mu - \\ 
-\partial _\mu \left( \partial _tK_t+\partial _xK_x+\partial _yK_y+\partial
_zK_z+\partial _\xi K_\xi +\partial _\kappa K_\kappa \right) =0\mbox{.}
\end{array}
\]

If

\[
\partial _tK_t+\partial _xK_x+\partial _yK_y+\partial _zK_z+\partial _\xi
K_\xi +\partial _\kappa K_\kappa =0 
\]

then

\[
\partial _t\partial _tK_\mu +\partial _x\partial _xK_\mu +\partial
_y\partial _yK_\mu +\partial _z\partial _zK_\mu =-\left( \partial _\xi
\partial _\xi K_\mu +\partial _\kappa \partial _\kappa K_\mu \right) \mbox{.}
\]

If $m_5$ and $m_4$ are a constant real numbers and for some\\function $W\mu
\left( t,x,y,z\right) $:

\[
K_\mu \left( t,x,y,z,\xi ,\kappa \right) =W\mu \left( t,x,y,z\right) \cdot
\exp \left( i\cdot \left( m_5\cdot \xi +m_4\cdot \kappa \right) \right) 
\]

then the particle, defined by the field $K_\mu $, has got the mass:

\[
m=\sqrt{m_5^2+m_4^2}\mbox{.} 
\]

It is a massive gauge boson.

\end{document}